\documentclass[aps,twocolumn, pra]{revtex4}

\usepackage{newlfont}
\usepackage{color}           % color to the text 
\usepackage{soul}            % strickthrough the text
\usepackage{amssymb}
\usepackage{amsfonts}
\usepackage{amsmath}
\usepackage{wasysym}
\usepackage{graphicx}
\usepackage{epsfig}
\usepackage{amsthm}
\usepackage{bm}

\begin{document}

%\title{Monogamy of Classical Capacity of Quantum States}
%\title{Monogamy of Quantum Dense Coding Capacity in arbitrary dimensions}

%\title{Strict Monogamy of Dense Coding Capacity of Quantum Channel}
%\title{Exclusion Principle and Monogamy for Dense Coding Capacity of Quantum Channel}
\title{Exclusion Principle for Quantum Dense Coding}

%\title{Conditions for Monogamy of Quantum Discord:\\ Monogamous Greenberger-Horne-Zeilinger versus Polygamous W states}
%\title{Conditions for Monogamy of Quantum Correlations:\\ Monogamous Greenberger-Horne-Zeilinger versus Polygamous W states}
%\title{Conditions for Monogamy of Quantum Correlations:\\ Polygamous W vs. Monogamous GHZ}
%\title{Conditions of Monogamy of Quantum Correlations: Polygamous W vs. Monogamous GHZ states}
%\title{Monogamy of Quantum Correlations as Detector of Genuine Tripartite Entangled States}

\author{R. Prabhu, Arun Kumar Pati, Aditi Sen(De), and Ujjwal Sen}

\affiliation{Harish-Chandra Research Institute, Chhatnag Road, Jhunsi, Allahabad 211 019, India}

\begin{abstract}
%Quantum correlations of quantum states are, generally, monogamous, while classical correlations of the same states are not so.

We show that the classical capacity of quantum states, as quantified by its ability to perform dense coding,
respects an exclusion principle, for arbitrary pure or mixed three-party states 
%of arbitrary 
in any dimension. 
%The monogamy behavior is strict, in the sense 
This states that no two bipartite states which are reduced states of a common tripartite quantum state 
can have simultaneous quantum advantage in dense
coding.
%In particular, we prove 
%%an exclusion principle for dense coding capacity which states 
%that if a sender has a quantum advantage in dense coding to one of the two receivers, 
%other receiver must necessarily have no quantum advantage.
%{\color{red}{\st{for}}} 
%{\color{green}{over}} 
The 
%strict monogamy 
exclusion principle is  robust against noise. 
%holds irrespective of whether the quantum channel carrying the post-encoding quantum 
%states from the sender to the receiver is noiseless or noisy. 
Such principle also holds for arbitrary number of parties.  
This exclusion principle is independent of the content and distribution of entanglement in the multipartite state. 
We also find  a strict monogamy relation for 
multi-port classical 
capacities of multi-party quantum states  in arbitrary dimensions.
In the scenario of two senders and a single receiver, we show that if two of them wish to send classical information to a single receiver 
%{\color{green}{receiver}} 
independently, then the corresponding dense coding capacities satisfy the monogamy relation, similar to the 
%{\color{green}{monogamy}} 
one for quantum correlations. 
\end{abstract}

\maketitle

\section{Introduction}

Quantum correlations  play an important role in quantum communication protocols \cite{comm-review}. Specifically,
entangled states have been used to transfer classical bits encoded in a quantum state beyond the classical limit (quantum dense coding) \cite{BW}, for transferring an unknown quantum state by using just two bits of classical 
communication (quantum teleportation) \cite{teleportation}, and for preparing a known quantum state at a remote location (remote state preparation) \cite{arun}.  
%All the protocols  can be shown to be advantageous over their classical counterpart. 
Such protocols were initially introduced for the case of a 
%when there is a 
single sender and a single receiver, and have also been experimentally realized \cite{exp}. However, for a fruitful application of such communication  schemes, 
it is of vital importance to consider an information transmission network that involves several senders and   receivers. 
%Therefore, investigations of  classical capacities or a quantum ones in a multipartite scanario  will be a step towarding achieving the goal of building communication network in 
%a quantum world.  

Study of correlations between separated physical systems is an important quantity in all areas of science. Such correlations can be classical as well as quantum. 
An important property of quantum correlations \cite{HHHH-RMP} in multipartite states
is that they tend to be ``monogamous'' in nature \cite{Ekert, Wootters, Wintermono, monogamyN}, in the sense that if two physical systems are highly quantum correlated, 
they cannot be correlated, individually or as a whole, with any third party. Monogamy of quantum correlations, therefore, restricts the sharability of quantum correlations between three or more parts of a quantum system. 
Classical correlations of 
 quantum states are certainly 
not monogamous, and an arbitrarily large number of physical systems can share the same amount of classical correlations with a single system.

In this paper, we address the question  whether there are restrictions on our ability to send classical information through quantum states used as quantum channels in a multipartite scenario (three or more parties). As noted above, there are no 
such restrictions on \emph{classical correlations} of quantum states. More precisely, for a three-party quantum state shared between Alice (\(A\)), Bob (\(B\)), and Charu (\(C\)), 
classical correlations between Alice and Bob, and between Alice and Charu can be both maximal. However, we
%, even when the senders and receiver share a quantum state. 
 show here that the \emph{classical capacity}, as quantified by the dense coding capacity, of an arbitrary (pure or mixed) three-party quantum state of arbitrary dimensions satisfies 
a strict monogamy relation that can be viewed as an exclusion principle:
If Alice has a quantum advantage in transferring classical information to Bob, she must necessarily have no quantum advantage in transferring the same to Charu. 
%$. Specifically, we show that if a quantum state is shared between three parties \(A\), \(B\) and \(C\),  and if 
%\(A\) decides to transfer classical information by using the shared reduced state to \(B\) and \(C\) independently, the quantum advatange is possible only either in the protocol of 
%\(A\) and \(B\) or in \(A\) and \(C\).    
This result 
%We show that this 
is independent of 
whether the quantum channel by which the quantum state of the sender is sent to the receiver in a dense coding protocol is
 noiseless or noisy. 
Note that this is stricter than the monogamy of quantum correlations (of quantum states): There exists quantum states for which Alice can have quantum correlations with 
Bob, and quantum correlations with Charu, i.e., the \(AB\) and the 
\(AC\) reduced quantum densities can both be quantum correlated, an example being the well-known three-party \(W\) state \cite{Wstate}.
We go on to show that the exclusion principle holds for an arbitrary  number of parties having an arbitrary amount of entanglement.

% We show this for an arbitrary (pure or mixed) three-party state of arbitrary dimensions.

%To our knowledge, the monogamous nature of two-party physical quantities have previously been obtained only for quantum correlations, and 

%only for multi-qubit systems. 

%Here we obtain a monogamy relation for channel capacity, and we

%obtain it for arbitrary dimensions.

%Remaining still w
Within the realm of tripartite states, we connect the monogamy of dense coding capacity to the monogamy relations known for quantum correlations. 
In particular, in the scenario of two senders and a single receiver, we show that if Bob and Charu wish to send classical information 
to Alice, then the corresponding dense coding capacities obeys the monogamy relation in the same spirit as for
quantum correlations. 
We subsequently generalize the monogamy relation to a multi-port scenario, involving multi-port channel capacities of 
multi-party (more than three-party) quantum states. 
%We show that in the case of several senders and a single receiver, the capacities are bounded above by the maximum entropy 
%of the senders. As a corollary, we also obtain a sufficient condition to satisfy monogamy of classical capacity for a pure quantum 
%tripartite state. 
We also establish
a relation between the sum of the capacities of dense coding in the \(AB\) and \(AC\) channels with the corresponding entanglements of formation \cite{EoF}, as well as their quantum discords \cite{discord}. 
This provides lower bounds to the sum of the capacities, 
complementary to the upper bounds obtained in the monogamy relations.

The paper is organized as follows. For completeness, we begin with a discussion of the quantum  dense coding capacity in Sec. II.
 Next, in Sec. III, we first prove the exclusion principle for dense coding capacity in the tripartite scenario. It holds for both the 
noiseless and noisy cases. We subsequently consider, in Sec. IV,  the multi-sender single-receiver scenario, and find a monogamy relation in that case. 
The case of multi-port channel capacities is considered in Sec. V and it is found that it also satisfies a strict monogamy.
% deals with such monogamy relation for a higher number of parties. 
We present a conclusion in Sec. VI.

\section{Quantum dense coding capacity}

Quantum dense coding is a quantum communication protocol that uses a shared quantum state between two distant observers, and a noiseless quantum channel \cite{dharma-versus-giraffe} to send classical information beyond the classical capacity of 
the quantum channel \cite{BW}.
Let 
%us call 
the observers, Alice and Bob, share the quantum state \(\varrho_{AB}\). Alice wishes to use this quantum state as a channel 
for sending classical information to Bob. Let the Hilbert space
which are in possession of Alice and Bob, and which supports the quantum state \(\varrho_{AB}\), be \({\cal H}_A \otimes {\cal H}_B\). 
%The suffixes \(A\) and \(B\) point to 
%Alice and Bob respectively. 
%The protocol is called quantum dense 
%coding, and runs as follows. 
Suppose that Alice receives a classical message \(i\), which is known to happen with probability \(p_i\). She encodes this 
classical message in a unitary operator \(U_i\) on the Hilbert space \({\cal H}_A\), 
and applies it to her part of \(\varrho_{AB}\) to obtain \(\varrho_{AB}^i = U_i\otimes \openone_B \varrho_{AB} U_i^\dagger \otimes \openone_B\), where 
\(\openone_B\) is the identity operator on the Hilbert space \({\cal H}_B\). She then sends her part of \(\varrho_{AB}^i\), through a noiseless quantum channel \cite{dharma-versus-giraffe} between Alice and Bob that 
can noiselessly transfer \(d_A\)-dimensional quantum states, to Bob. Here, \(d_A = \dim {\cal H}_A\). 
After this, 
%Alice's job is
% completed, 
Bob is in possession of the quantum ensemble
\(\{p_i,\varrho_{AB}^i\}\), and his task is to perform a quantum measurement on this ensemble so as to obtain as much information as possible about the classical index \(i\).

After the quantum measurement by Bob, suppose that the post-measurement quantum ensemble is \(\{p_{i|m},\varrho_{AB}^{i|m}\}_i\), and also suppose that this ensemble
 appears with probability \(q_m\). The amount of classical information
gained by Bob due to his measurement can be quantified by the mutual information \cite{eitaCoverThomas} between the index \(i\) and the measurement index \(m\), and is given by 
\begin{equation}
 I(i:m) = H(\{p_i\}) - \sum_m q_m H(\{p_{i|m}\}_i)
\end{equation}
bits,
where \(H(\cdot)\) denotes the Shannon entropy of the probability distribution in its argument. The unit of mutual information is taken here to be ``bits'', a result of the fact that we are using the logarithms with base 2 in this paper, for 
both Shannon and von Neumann entropy. Henceforth, all the entropic quantities are defined in bits.

Now Bob has to perform a measurement that maximizes his information gain, and this information is the ``accessible information''
defined as
\begin{equation}
 I_{acc}(\{p_i,\varrho_{AB}^i\}) = \max I(i:m),
\end{equation}
 where the maximization is over all measurement strategies that Bob is able to implement on his ensemble.

This maximization turns out to be hard to implement. However, an useful upper bound,  called the Holevo bound \cite{Holevo, see-also-Holevo}, exists, and is given by 
\begin{equation}
 \chi(\{p_i,\varrho_{AB}^i\}) = S(\overline{\varrho}_{AB}) - \sum_ip_iS(\varrho^i_{AB}),
\end{equation}
where \(S(\cdot)\) is the von Neumann entropy of the quantum state in its argument, and \(\overline{\varrho}\) is the average ensemble state \(\sum_ip_i \varrho^i_{AB}\). 
This quantity is asymptotically achievable \cite{Holevo-ananta}, and therefore the following quantity is termed the dense coding capacity of the quantum state \(\varrho_{AB}\):
\begin{equation}
{\cal C}(\varrho_{AB}) = \max_{\{p_i, U_i\}} \chi (\{p_i,\varrho_{AB}^i\}).
\end{equation}

It is possible to perform this optimization \cite{dcgeneral, dcamader}, and one obtains 
\begin{equation}
\label{eq:capdefi}
 {\cal C}_{AB} \equiv  {\cal C}(\varrho_{AB}) = \log_2 d_A + S(\varrho_B) - S(\varrho_{AB}),
\end{equation}
 where \(\varrho_B = \mbox{tr}_A[\varrho_{AB}]\).
It is to be noted that the conditional entropy \(S(\varrho_{AB}) - S(\varrho_B)\) can be of both signs.
If it is positive,  one may not use the shared quantum state, but use the noiseless quantum channel to transfer 
\(\log_2 d_A\) bits of classical information. In case the conditional entropy is negative, Alice will be able to use the shared quantum state to send classical information, 
beyond the ``classical limit'' of \(\log_2 d_A\) bits, to Bob. We term this as 
a ``quantum advantage'' for Alice in sending classical information to Bob.
% and this is possible due to the existence of entanglement between A and B. 
So in general, 
the dense coding capacity is given by $\overline{{\cal C}}_{AB}\equiv\overline{{\cal C}}(\varrho_{AB})=\mbox{max}[\log_2 d_A,\, {\cal C}(\varrho_{AB})]$,
and we term \({\cal C}_{AB}\) as the quantum part of the dense coding capacity.

\section{Exclusion Principle for dense coding capacity for three-party states}

In this section, we will begin by presenting the exclusion principle for an arbitrary (pure or mixed) three-party quantum state of arbitrary dimensions.

\noindent {\bf Theorem 1:} (``Exclusion Principle'') 
\emph{Given an arbitrary (pure or mixed) three-particle quantum state \(\varrho_{ABC}\), no two bipartite states shared with 
any one of the parties
can have a quantum advantage in dense coding capacity simultaneously.}
%For an arbitrary (pure or mixed) three-party quantum state $\varrho_{ABC}$, shared between \(A\), \(B\), and \(C\), in arbitrary dimensions, 
%the sum of the dense coding capacities of 
%the reduced states $\varrho_{AB}$ and $\varrho_{AC}$ 
%is bounded above by $2 \log_2 d_A$, where $d_A= \dim {\cal H}_A$ is the dimension of $A$.}\\
%any two reduced states \
%cannot possess quantum advantages in dense coding simultaneously.}
\\

%Given an arbitrary (pure or mixed) three-particle quantum state \(\varrho_{ABC}\), no two bipartite states shared with 
%any one of the parties
%can have a quantum advantage in dense coding capacity simultaneously.

\noindent \texttt{Proof.} Let  us assume the contrary and suppose that both \(\varrho_{AB}\) and  \(\varrho_{AC}\) have quantum advantages in dense coding, where 
%Then 
%$\varrho_{ABC}$ be shared state between \(A\), \(B\) and \(C\) and let 
%${\cal C}_{AB}$ 
%and ${\cal C}_{AC}$ 
%be the dense coding capacity when Alice wishes to send classical information to Bob by using the reduced quantum density matrix 
\(\varrho_{AB} = \mbox{tr}_{C}[\varrho_{ABC}]\) and \(\varrho_{AC} = \mbox{tr}_{B}[\varrho_{ABC}]\). 
%defined in  Eq. (\ref{eq:capdefi}). 
%${\cal C}_{AC}$ is similarly defined. Also, \(\varrho_C = \mbox{tr}_{AB} [\varrho_{ABC}]\), and similarly for \(\varrho_A\) and \(\varrho_B\).
Then, we have
\begin{eqnarray}
\overline{{\cal C}}_{AB} + \overline{{\cal C}}_{AC} = {\cal C}_{AB} + {\cal C}_{AC} \phantom{aaaaaaaaaa}\nonumber \\
= 2\log_2 d_A + S(\varrho_B) + S(\varrho_C) - S(\varrho_{AB}) - S(\varrho_{AC}),
\end{eqnarray}
with \(\varrho_C = \mbox{tr}_{AB} [\varrho_{ABC}]\), and similarly for \(\varrho_A\) and \(\varrho_B\).
Strong subadditivity of von Neumann entropy \cite{ekhane-NC} for the tripartite system  between \(A\), \(B\), and \(C\) implies that  
\begin{equation}
 S(\varrho_B) + S(\varrho_C) - S(\varrho_{AB}) - S(\varrho_{AC}) \leq 0.
\end{equation}
Therefore,  we get
\begin{equation}
\label{eq:main}
\overline{{\cal C}}_{AB} + \overline{{\cal C}}_{AC}  \leq 2\log_2 d_A.
\end{equation}
Equality sign will be satisfied by all pure three-party states. 

But, if both \(\varrho_{AB}\) and  \(\varrho_{AC}\) have quantum advantages, then by definition of the dense coding capacity, \(\overline{{\cal C}}_{AB} + \overline{{\cal C}}_{AC}\) 
must be strictly greater than \(2\log_2 d_A\), contradicting our assumption. 
\hfill $\blacksquare$

\noindent \emph{Remark:} Note that Theorem 1 can also be interpreted as a strict monogamy relation of the dense coding capacity:
%
%: Given a tripartite state \(\varrho_{ABC}\), no two bipartite states shared with 
%any one of the parties
%can have a quantum advantage in dense coding capacity simultaneously. 
%In other words,  
If Alice has a quantum advantage in sending classical information to Bob (i.e. if \({\cal C}_{AB} > \log_2 d_A\)), then 
Alice cannot have a quantum advantage with Charu (i.e., \({\cal C}_{AC}\) must necessarily be strictly less than \(\log_2 d_A\)), 
so that Alice will be forced to send classical information at the classical limit rate to Charu which is equal to \(\log_2 d_A\).

%We observe that the Theorem 1 is useful to conclude about the quantumness of the channels which are required for the dense coding in multipartite scenario. We study this observation by considering the following two consecutive corollaries.
%\\
\noindent {\em Corollary 1:} In a tripartite quantum state $\varrho_{ABC}$, if $\varrho_{AB}$ and $\varrho_{AC}$ are two reduced quantum states 
through which Alice wants to send classical information to Bob and Charu, then the sum of the dense coding capacities of the reduced states $\varrho_{AB}$ and 
$\varrho_{AC}$ is bounded above by $3\log_2 d_A$. The bound can be saturated.
\\

\noindent \texttt{Proof.} 
%The dense coding capacity for the channels $\varrho_{AB}$ and $\varrho_{AC}$, when quantumness is not known, can be written as
%\begin{eqnarray}
%\overline{{\cal C}}_{AB} &=& \mbox{max}[\log_2 d_A,\, \log_2 d_A + S(\varrho_B) - S(\varrho_{AB})]\,\, \mbox{and} \nonumber\\
%\overline{{\cal C}}_{AC} &=& \mbox{max}[\log_2 d_A,\, \log_2 d_A + S(\varrho_C) - S(\varrho_{AC})], \nonumber
%\end{eqnarray}
%respectively. 
From Theorem 1, it follows that the two channels cannot have  quantum advantages simultaneously. Hence there are two possibilities -- 
(i) both of them are classical, which implies $\overline{{\cal C}}_{AB} + \overline{{\cal C}}_{AC} =2\log_2 d_A$, and (ii) 
%It is clear that ${\cal C}_{AB} + {\cal C}_{AC}$ will reach maximum, only
one of the channels is classical and the other quantum (i.e. has a quantum advantage). 
In the case (ii), without loss of generality, we assume that the \(AB\) channel is quantum. 
% we can get optimal capacity when (by choosing $\varrho_{AB}$ 
%as quantum and $\varrho_{AC}$ as classical channel)
Therefore,
\begin{eqnarray}
\overline{{\cal C}}_{AB} + \overline{{\cal C}}_{AC} = 2\log_2 d_A + S(\varrho_B) - S(\varrho_{AB}).
\label{optimalcap}
\end{eqnarray}
%Recently, Carlen and Lieb \cite{CarlLieb} have shown that the entanglement of formation of a bipartite state is related to the conditional entropies 
%$S(\varrho_{A|B})$ and $S(\varrho_{B|A})$ as  follows:
The strong subadditivity of von Neumann entropy implies that 
\begin{equation}
S(\varrho_B) - S(\varrho_{AB}) \leq S(\varrho_{AC}) - S(\varrho_{C}).
%E_f({\varrho_{AB}})=\mbox{max}[S(\varrho_{A|B}),\, S(\varrho_{B|A}),\, 0].\nonumber
\end{equation}
On the other hand, the nonnegativity of quantum mutual information implies that 
\begin{equation}
 S(\varrho_{AC}) - S(\varrho_{C}) \leq S(\varrho_A),
\end{equation}
so that we have
\begin{equation}
S(\varrho_B) - S(\varrho_{AB}) \leq S(\varrho_{A}) \leq \log_2 d_A.
%E_f({\varrho_{AB}})=\mbox{max}[S(\varrho_{A|B}),\, S(\varrho_{B|A}),\, 0].\nonumber
\end{equation}
Using this relation in Eq. (\ref{optimalcap}), we obtain
\begin{eqnarray}
\overline{{\cal C}}_{AB} + \overline{{\cal C}}_{AC} \leq 3\log_2 d_A
% + E_f(\varrho_{AB}) \nonumber\\
%&\leq& 2\log_2 d_A + \log_2 d_A \nonumber\\
%&\leq& 3\log_2 d_A. \nonumber
\end{eqnarray}
The proof follows by combining the cases (i) and (ii). \hfill \(\blacksquare\)

%Thus we have obtained the upper bound for capacities when one of the channel is classical and other being quantum. 
We now generalize our findings to states of more than three parties.
% for the 
%multiparty senders and single receiver.

\noindent {\bf Theorem 2:} \emph{For an arbitrary  (pure or mixed) multiparty state $\varrho_{AB_1B_2\ldots B_N}$, shared between $(N+1)$ parties, in arbitrary dimensions,
only at most a single reduced density matrix among \(\varrho_{AB_i}\) \((i=1,2,\ldots,N)\) can have quantum advantage in dense coding.}
% the sum of the dense coding capacities of the reduced states $\varrho_{AB_1},\, \varrho_{AB_2},\, \ldots\, \varrho_{AB_N}$ is bounded above by $N\log_2 d_A$.}
\\

\noindent \texttt{Proof.} 
Suppose, if possible, that \(\varrho_{AB_{k_1}}\) and \(\varrho_{AB_{k_2}}\) have quantum advantages in dense coding. However, in that case, the reduced states 
%tripartite quantum state
\(\varrho_{AB_{k_1}}\) and \(\varrho_{AB_{k_2}}\) of the tripartite quantum state 
\(\varrho_{AB_{k_1}B_{k_2}}\)
%the reduced state of  \(A\), \(B_{k_1}\), and \(B_{k_2}\),
violates Theorem 1.
%
%
%From Theorem 1, for any two channels we have 
%
%
%\begin{eqnarray}
%{\cal C}_{AB_1} + {\cal C}_{AB_2} &\leq& 2\log_2 d_A \nonumber\\
%{\cal C}_{AB_2} + {\cal C}_{AB_3} &\leq& 2\log_2 d_A \nonumber\\
%&\vdots& \nonumber\\
%{\cal C}_{AB_1} + {\cal C}_{AB_N} &\leq& 2\log_2 d_A. \nonumber
%\end{eqnarray}
%These equations will add up to give us
%\begin{equation}
%\sum_{i=1}^{N} {\cal C}_{AB_i} \leq N\log_2 d_A. \nonumber
%\end{equation}
%Hence the proof. 
\hfill \(\blacksquare\)

%\noindent {\bf Corollary 2:}

%\emph{In other words, if a $N$-partite quantum state $\varrho_{AB_1B_2\ldots B_N}$, if $\varrho_{AB_1},\, \varrho_{AB_2},\, \ldots\, \varrho_{AB_N}$ 
%are the reduced quantum states through which one sender wants to send classical information to many receivers, then no \((N-1)\)-pair 
%can not possess quantum advantage simultaneously.}
%This is due to teh fact that if more than one pair can have quantum advantage, then Theorem 1 will be violated.
%\\
%\\
%\\

\noindent{\em Corollary 2:} In an $(N+1)$-party quantum state $\varrho_{AB_1B_2\ldots B_N}$, the sum of the dense coding capacities 
in the cases
%reduced two-party density matrices,
 where \(A\) is the sender and \(B_i, i= 1, 2, \ldots, N\) are the receivers,  is bounded above
\( (N+1) \log_2 d_A\). 
\\

%\noindent \texttt{Proof.} The proof is straight forward from the Corollary 1, which gives us the following relation for multiparty scenario
%\begin{equation}
%\overline{{\cal C}}_{AB_1} + \overline{{\cal C}}_{AB_2} + \cdots + \overline{{\cal C}}_{AB_N} \leq (N+1) \log_2 d_A.
%\end{equation}

Until now, we have considered the situation where the quantum channel, carrying Alice's part of the states 
%\(\varrho_{AB}^i\) (or \(\varrho_{AC}^i\)) 
to the receivers
%Bob (or Charu),
 as noiseless. It turns out that the exclusion principle holds also in a more general scenario, 
when the aforementioned quantum channel is noisy. 
%We will consider the case when the quantum channel is a 
%
%Let us now consider the case when the quantum channel, through which Alice's part of the ensemble states \(\varrho_{AB}^i\) are sent to Bob, is noisy.
% noisy channel, i.e. consider the case when after unitary encoding, the ensembles \(\{p_i, \varrho_i\}\) can be sent via noisy channel.
 %We will for example consider 
%covariant noisy channel \cite{ekhane-covariant-noisy}, denoted as \(\Lambda\), and defined as one  which commutes with a complete set of orthogonal unitary operators.
%A similar relation can also be proven corresponding to Theorem 2.
%
% and hence
%\(\Lambda (U_i \varrho U_i^{\dagger}) = U_i  \Lambda(\varrho) U_i^{\dagger}\). 
%This is true for a complete set of orthogonal unitary operators \(\{U_i\}\) which satisfy the trace rule \((1/d) \sum_i U_i A U_i^{\dagger} = \mbox{tr}(A) I\) for any operator \(A\).
%Below we prove the exclusion principle for the dense coding capacity for noisy channels.
%\noindent {\bf Remark:} (``Exclusion Principle for Noisy Channel'') For an arbitrary (pure or mixed) three-party quantum state $\varrho_{ABC}$, 
%%shared between \(A\), \(B\), and \(C\), in arbitrary dimensions, 
%an upper bound of the sum of 
%%the reduced states $\varrho_{AB}$ and $\varrho_{AC}$ 
%is bounded above by $2 \log_2 d_A$, where $d_A= \dim {\cal H}_A$ is the dimension of $A$.}\\
%any two reduced states \
%%cannot possess quantum advantages in dense coding simultaneously,
 %%even when a noisy channel \(\Lambda\) is acting on the quantum states sent from the 
%%sender to the receivers. 
This is due to the fact that the capacities  will be non-increasing in the presence of noise. Therefore, the upper bound, obtained in Theorems 1 and 2 also hold for any noisy channel.  Henceforth, we consider only noiseless channels.

\section{Reveiver Monogamy for dense coding capacities}
\label{bhalo}

For quantum correlations, to check for the status of monogamy for a particular measure, one usually considers 
inequalities where the sum of the Alice-Bob and Alice-Charu quantum correlations is compared with that share by Alice with 
the Bob-Charu pair. We now consider the status of such relations, when taken over to the case of dense coding capacities. We begin with the case where two senders (Bob and Charu) wish to send information to Alice by using 
the three-party quantum state \(\rho_{ABC}\). Again the state can be either pure or mixed, and in arbitrary dimensions.

%Next, we prove a theorem that shows that the dense coding capacites from two senders to a single reciver obey actual mpnogamy relation (from the receiver's perspective).

Let \({\cal C}_{BA}\) be the quantum part of the 
dense coding capacity when Bob wants to send classical information to Alice by using the reduced density state \(\varrho_{AB}\). 
Let \({\cal C}_{CA}\) be similarly defined. Let 
\({\cal C}_{BC:A}\) be the quantum part of the dense coding capacity when Bob and Charu sends classical information to Alice by using the quantum state \(\varrho_{ABC}\).
%, when Bob and Charu are together.

\noindent {\bf Theorem 3:} (``Receiver Monogamy'') \emph{For an arbitrary tripartite pure or mixed 
quantum state $\varrho_{ABC}$, shared between \(A\), \(B\), and \(C\) in arbitrary dimensions, the dense coding capacities are such that the monogamy} 
$${\cal C}_{BA} + {\cal C}_{CA}\leq  {\cal C}_{BC:A},$$ \emph{is satisfied, even when \(B\) and \(C\) are far apart.}\\
%, where $B, C$ are senders and $A$ is the receiver.\\

\noindent \texttt{Proof:} 
%Let  $\varrho_{ABC}$ be shared state between \(A\), \(B\), and \(C\) and let us define 
We have 
${\cal C}_{BA}  = \log_2 d_B + S(\varrho_A) - S(\varrho_{AB})$ and 
and ${\cal C}_{CA} =  \log_2 d_C + S(\varrho_A) - S(\varrho_{AC})$, where \(d_B\) and \(d_C\) are the dimensions of the Hilbert spaces in possession of Bob and Charu respectively.
Now, using strong subadditivity of von Neumann entropy \cite{ekhane-NC} for a tripartite system  between \(A\), \(B\), and \(C\),  we have
\begin{equation}
 S(\varrho_A) -  S(\varrho_{AB}) + S(\varrho_A) - S(\varrho_{AC}) \leq S(\varrho_A) - S(\varrho_{ABC}) 
\end{equation}
so that 
\begin{equation}
{\cal C}_{BA} + {\cal C}_{CA} \leq \log_2 (d_B d_C) + S(\varrho_A) - S(\varrho_{ABC}).
\label{eq:main1}
\end{equation}
However, the quantum part of the dense coding capacity of $BC$ to $A$ is ${\cal C}_{BC:A}  = \log_2 (d_B d_C)  + S(\varrho_A) - S(\varrho_{ABC})$.
Note here that for Bob and Charu to attain a dense coding capacity of \(\log_2 (d_B d_C)  + S(\varrho_A) - S(\varrho_{ABC})\) for sending classical information to Alice, it is not necessary for Bob and Charu to come together, as the dense coding capacity is attained by local encodings \cite{dcamader} (cf. \cite{Michal-general}).
Hence, the theorem. \hfill $\blacksquare$

We now consider the relation stated in Theorem 3 
in the situation when Alice is the sender, instead of being the receiver of the dense coding channels. Let us therefore compare the sum of the quantities 
${\cal C}_{AB}$ and \({\cal C}_{AC}\) with the quantum part of the dense coding capacity, \({\cal C}_{A:BC}\), 
when Alice wants to send classical information to Bob and Charu (who are together) by using a shared quantum state between the three parties.

\noindent \emph{Corollary 3:} A tripartite pure state \(|\psi_{ABC}\rangle\) satisfies the relation 
${\cal C}_{AB} + {\cal C}_{AC}\leq  {\cal C}_{A:BC}$, only if it possesses maximal entanglement between Alice and the Bob-Charu pair.\\
%, i.e. if the local von Neumann
%entropy of Alice,  \(S(\varrho_A)\), is maximal. 

\noindent \texttt{Proof.} 
For a pure three-party state $|\psi_{ABC}\rangle$, Theorem 1 implies that  
${\cal C}_{AB} + {\cal C}_{AC}= 2\log_2 d_A$. 
The quantum part of the dense coding capacity when $A$ is sending to the $BC$ pair (with the latter being together) is given by 
${\cal C}_{A:BC}=\log_2 d_A + S(\varrho_{BC})$. Therefore, the relation ${\cal C}_{AB} + {\cal C}_{AC}\leq  {\cal C}_{A:BC}$ for the quantum parts of the capacities
 reduces to
 $\log_2 d_ A \leq S(\varrho_{BC}) = S(\varrho_{A})$. But the entropy of a system cannot be more than the logarithm of its 
 dimension, i.e., $S(\varrho_A) \geq \log_2 d_A$. Therefore,  $\log_2 d_A = S(\varrho_{BC}) = S(\varrho_{A})$. Also, maximal local entropy for a pure 
bipartite state implies that it is maximally entangled. 
 Therefore, the 
 entanglement in the $A:BC$ bi-partition has to be maximum, if the dense coding capacities satisfies the relation in the premise of the theorem. 
% for pure tripartite states. In case of three qubits, the monogamy is satisfied only when  one ebit of entanglement is 
% present between $A$ and $BC$ partitions.
\hfill $\blacksquare$

Note that in the case of three-qubit pure states, the relation ${\cal C}_{AB} + {\cal C}_{AC}\leq  {\cal C}_{A:BC}$ is satisfied only when the state has  one ebit of entanglement in its \(A:BC\) partition. 
% present between $A$ and $BC$ partitions.

\noindent \emph{Corollary 4:} If a tripartite pure or mixed state \(\varrho_{ABC}\) satisfies the monogamy relation ${\cal C}_{AB} + {\cal C}_{AC}\leq  {\cal C}_{A:BC}$,
% of dense coding capacities, 
then the state should satisfy the following inequality:
\begin{equation}
\label{eq:mixedP}
\log_2 d_A- S(\varrho_{A}) \leq \sum_{i=A,B,C} S(\varrho_{i}) -S(\varrho_{ABC}). 
\end{equation}
\\

\noindent \texttt{Proof.} For an arbitrary tripartite pure or mixed state \(\varrho_{ABC}\),  
the monogamy relation ${\cal C}_{AB} + {\cal C}_{AC}\leq  {\cal C}_{A:BC}$  can be written by using Eq. (\ref{eq:capdefi}) as
\begin{eqnarray}
\log_2 d_A + S(\varrho_{B}) + S(\varrho_{C}) \leq S(\varrho_{AB}) + S(\varrho_{BC}) \nonumber\\
+ S(\varrho_{AC})- S(\varrho_{ABC}).
\end{eqnarray}
Using the subadditivity of entropy \cite{ekhane-NC}, i.e., $S(\varrho_{AB}) \leq S(\varrho_{A}) + S(\varrho_{B})$, and after rearrangement, we obtain the stated sufficient condition.
%for a arbitrary mixed state to satisfy the monogamy of dense coding capacity, given in Eq. (\ref{eq:mixedP}). 
\hfill $\blacksquare$

%$$\log d_A-S_A \leq  S(\varrho_{i}) -S(\varrho_{ABC})$$.

We will now derive a lower bound on the sum, ${\cal C}_{AB} + {\cal C}_{AC}$, of the quantum parts of the 
capacities, in terms of measures of quantum correlations. Let the 
entanglements of formation  \cite{EoF} between Alice and Bob, and between Alice and Charu  be $E_{AB}$ and $E_{AC}$ respectively.   Also, suppose that the quantum discords \cite{discord} between Alice and Bob, and between 
Alice and Charu are $D_{AB}$ and $D_{AC}$ respectively.

\noindent \emph{Corollary 5:} The sum of the quantum parts of the capacities, ${\cal C}_{AB}$ and ${\cal C}_{AC}$, of a tripartite pure state $|\psi_{ABC}\rangle$ is bounded below by $D_{AB}+D_{AC} = E_{AB} + E_{AC}$. \\

\noindent \texttt{Proof.} In case of a pure tripartite state $|\psi_{ABC}\rangle$, Koashi and Winter \cite{monogamyN} have found a 
relation between the bipartite entanglement of formation  and bipartite quantum discord, which
reads
 $E_{AB}=D_{AC}+S(\varrho_{A|C}),$ 
where $S(\varrho_{A|C}) = S(\varrho_{AC})- S(\varrho_{C})$ is the conditional entropy. 
By using Eq. (\ref{eq:capdefi}), one obtains
 ${\cal C}_{AB} = D_{AB} - E_{AC} + \log_2 d_A$, 
and the quantum part of the capacity between $A$ and $C$ is
 ${\cal C}_{AC}=D_{AC}-E_{AB}+\log_2 d_A$. 
The sum of these two quantities will then give
\begin{equation}
\label{eq:kirchoff}
{\cal C}_{AB} + {\cal C}_{AC} = D_{AB} + D_{AC} - E_{AB}- E_{AC} + 2\log_2 d_A. 
\end{equation}
Moreover, the sum of the entanglements of formation of \(AB\) and \(AC\) are bounded above by \(2 \log_2 d_A\) \cite{EoF}, i.e. 
$E_{AB}+E_{AC} \leq 2\log_2 d_A$. 
This immediately implies that 
\begin{equation}
{\cal C}_{AB} + {\cal C}_{AC} \geq D_{AB}+D_{AC} = E_{AB} + E_{AC}.
\label{eq:capdisrel}
\end{equation}
To obtain the last equality, we use Theorem 1 in Eq. (\ref{eq:kirchoff}) which leads to $D_{AB} + D_{AC} -  E_{AB} - E_{AC}=0$.
\hfill $\blacksquare$

%Recently, it was shown that a state belonging to the  \(W\)-class \cite{Wstate} invariably violates the monogamy for quantum discord \cite{amaderdisco}, i.e., for these states, 
% $D_{AB} + D_{AC} \geq S(\varrho_{A})$.  Therefore, for the states from  the \(W\)-class, the sum \({\cal C}_{AB} + {\cal C}_{AC}\),     of dense coding capacities is bounded below by $S(\varrho_A)$, so that in conjunction with  Theorem 1 implies that 
%$$S(\varrho_{A}) \leq {\cal C}_{AB}+{\cal C}_{AC} = 2 \log_2 d_A.$$
%Since $S(\varrho_{A})$ is strictly positive for W-class states, the lower bound obtained above is nontrivial. 

\section{Monogamy of multi-port dense coding capacities}

In this section, we generalize the strict monogamy relations to an arbitrary number of parties for the case of multi-port capacities. 
Let us consider a situation where there are \(N\) observers, whom we call Alices (\(A_1\), \(A_2\), \(\ldots\), \(A_N\)), and who 
share an \(N\)-party quantum state \(\varrho_{A_1A_2 \ldots A_N}\). Let \({\cal C}_{A_1A_2\ldots A_{N-2}A_{N-1}}\) denote the quantum part of the
``distributed'' or ``multi-port'' dense coding capacity in the case when all Alices except \(A_{N-1}\) and \(A_{N}\)
are senders, and \(A_{N-1}\) is the receiver. Let \({\cal P}_{N-1}^N\) denote a periodic shift operator that takes \(N-1\) elements from the ordered 
periodic collection \(A_1 A_2 \ldots A_N\), so that 
\({\cal P}_{N-1}^N A_1A_2\ldots A_{N-2}A_{N-1} = A_2A_3\ldots A_{N-1}A_{N}\), \(({\cal P}_{N-1}^N)^2 A_1A_2\ldots A_{N-2}A_{N-1} = A_3A_4\ldots A_{N}A_{1}\), etc. 
Therefore, we can visualize the \(N\) Alices as situated on different points in a ring.
We suppose that they are ordered and we assume that the ordering has been performed in the clockwise direction. Any 
consecutive \(N-2\) of them are acting as senders, and they are trying to send classical information to the Alice who is situated just beside them in a clockwise direction.

%Here \(\overline{A_j}\) denotes the set of all senders except \(A_j\).

\noindent {\bf Theorem 4:} (``Strict Monogamy for Multi-port Capacities'') 
\emph{For an arbitrary pure or mixed quantum state \(\varrho_{A_1A_2 \ldots A_N}\) in arbitrary dimensions, 
the quantum parts of the distributed dense coding capacities satisfy the following strict monogamy relation:} 
\begin{eqnarray}
\label{eq:maingeneral}
\sum_{j=0}^{N-1} {\cal C}_{({\cal P}_{N-1}^N)^j A_1A_2\ldots A_{N-2}A_{N-1}}
\leq (N-2) \sum_{j=1}^N \log_2 d_{{A}_{j}}, \quad
\end{eqnarray}
\emph{where \(d_{A_j}\) is the dimension of the Hilbert space in possession of \(A_j\).}\\
%each of the capacities involve \(N-1\) parties, among which first \(N-2\) parties are the sender and a rest is the receiver.  

\noindent \texttt{Proof.} 
%Let us consider a \((N-1)\)-party quantum state \(\varrho_{A_1A_2 \ldots A_{N-1}}\). 
%Assume that  first \(N-2\) parties, \(A_1, A_2, \ldots, A_{N-2}\) serve as senders and \(A_{N-1}\) is a single receiver.  In this situation, 
The quantum part of the distributed dense coding capacity \({\cal C}_{A_1A_2\ldots A_{N-2}A_{N-1}}\) 
 is given by \cite{dcamader}
\begin{eqnarray}
\label{eq:capgen}
{\cal C}_{A_1A_2\ldots A_{N-2}A_{N-1}}  = \sum_{i=1}^{N-2} \log_2 d_{A_i} \phantom{aaaaaaaaaaaaaa} \nonumber \\ + S(\varrho_{A_{N-1}}) - S(\varrho_{A_1A_2\ldots A_{N-2}A_{N-1}}),
\end{eqnarray}
 in which the senders are allowed to perform unitary encoding.  Here, \(\varrho_{A_{N-1}} = \mbox{tr}_{A_1A_2\ldots A_{N-2}A_N} \varrho_{A_1 A_2 \ldots A_N}\) and 
\(\varrho_{A_1A_2\ldots A_{N-2}A_{N-1}} = \mbox{tr}_{A_N}\varrho_{A_1A_2 \ldots A_N}\). Below, the local densities are defined similarly.
  Using  Eq. (\ref{eq:capgen}), we have
%, the monogamy relation in Eq. (\ref{eq:maingeneral}) reduces to
\begin{eqnarray}
\sum_{j=0}^{N-1} {\cal C}_{({\cal P}_{N-1}^N)^j A_1A_2\ldots A_{N-2}A_{N-1}}
%\sum_{j=1}^N {\cal C}_{\overline{A}_jB}
% {\cal C}_{A_1A_2 \ldots A_{N-1}} + {\cal C}_{A_1 A_3 \ldots A_N} + \ldots  {\cal C}_{A_1A_2 \ldots A_{N} A_{N-2}}\\
= (N-2)\sum_{j=1}^N \log_2 d_{A_j} \nonumber\\
+ \sum_{j=1}^{N} S(\varrho_{A_j}) - \sum_{j=0}^{N-1} S(\varrho_{({\cal P}_{N-1}^N)^j A_1A_2\ldots A_{N-2}A_{N-1}}), 
\label{eq:proof}
\end{eqnarray}
%where the last sum consists of all the combination of \((N-1)\)-parties except one in which \(A_1\) has to serve as a receiver. 
To prove the nonpositivity of the last line in the above equation (Eq. (\ref{eq:proof})), we will need the strong subadditivity of von Neumann entropy involving \(N\) parties, which we now establish, for completeness. 
We have 
%Now,
 \begin{eqnarray}
\sum_{j=1}^{N} S(\varrho_{A_j}) - \sum_{j=0}^{N-1} S(\varrho_{({\cal P}_{N-1}^N)^j A_1A_2\ldots A_{N-2}A_{N-1}}) \nonumber \\
%N S(\varrho_B) - \sum_{j=1}^N S(\varrho_{\overline{A}_j B}) = - \sum_{j=1}^N S(\varrho_{B|\overline{A}_j}) \\ \nonumber
= - \sum S(\varrho_{R_j|({\cal P}_{N-1}^N)^j A_1A_2\ldots A_{N-2}A_{N-1}}) \nonumber \\
\equiv {\cal Q}(\varrho_{A_1A_2 \ldots A_N}), 
 \end{eqnarray}
where \(R_j\) is the observer which is left out from the \(N\) Alices in the collection \(({\cal P}_{N-1}^N)^j A_1A_2\ldots A_{N-2}A_{N-1}\),  and
\(S(\varrho_{R_j|({\cal P}_{N-1}^N)^j A_1A_2\ldots A_{N-2}A_{N-1}})\) is the conditional entropy defined as 
\(S(\varrho_{({\cal P}_{N-1}^N)^j A_1A_2\ldots A_{N-2}A_{N-1}}) - S(\varrho_{R_j})\).
%\(S(\varrho_{\overline{A}_j B}) - S(\varrho_B)\).
%with \(j\ne i\). 
Since the conditional entropies are convex, \( {\cal Q}(\varrho_{A_1A_2 \ldots A_N})\) is also a convex function. Moreover \(\varrho_{A_1A_2 \ldots A_N }\) can be written in a 
spectral decomposition as  \( \sum p_k |K\rangle \langle K|\). So,   \( {\cal Q}(\varrho_{A_1A_2 \ldots A_N }) \leq \sum p_k {\cal Q}(|K\rangle \langle K|)\).  However, 
\( {\cal Q}(\varrho_{A_1A_2 \ldots A_N}) =0\) for 
pure states.
%, since eigenvalues of the local density matrices of a pure state are the same. 
%is equal to the eigenvalues of the rest. 
Therefore, 
\[ \sum_{j=1}^{N} S(\varrho_{A_j}) - \sum_{j=0}^{N-1} S(\varrho_{({\cal P}_{N-1}^N)^j A_1A_2\ldots A_{N-2}A_{N-1}}) \leq 0. \]
 Hence the theorem. 
%-port dense coding capacity follows  is satisfied for arbitrary statee with arbitrary number of parties in arbitrary dimensions.  
\hfill $\blacksquare$

\noindent \emph{Remark:} Theorem 4 implies that not all groups of \(N-2\) senders can get a quantum advantage in sending classical information to the corresponding receiver. 
% Therefore, not all $N$ senders can get quantum advantage in sending classical information to Bob. 
They must respect the monogamy relation, given in Eq. (\ref{eq:maingeneral}). There are \(N\) such sender groups and at most \(N-1\) sender groups can have quantum advantages. In other words, if \(N-1\) sender groups have quantum advantages, the \(N\)th 
sender group must necessarily have no quantum advantage in sending classical information to their intended receiver. In this sense, the monogamy for multi-port capacities is again strict. 

\section{Conclusion}

Usually, \emph{quantum} correlations are expected to obey monogamy. However, in this paper, we have found that \emph{classical} capacity of a quantum 
channel obeys an extreme form of monogamy, which we refer as an exclusion principle. 
Specifically, we have shown that in a tripartite scenario, if Alice, Bob, and Charu share an arbitrary tripartite (pure or mixed) state in arbitrary dimensions,
and Alice wishes to send classical information, 
encoded in a quantum state, to Bob and Charu independently, then quantum protocols can give  advantage over classical   
ones either in the  Alice-Bob protocol or in the Alice-Charu protocol. This is also true for an arbitrary number of parties in arbitrary dimensions.
This exclusion principle is independent of the shared entanglement
between the parties. 
%We have also proved that this
The principle also holds in the case when the quantum channel carrying the post-encoding quantum states from the sender to the receiver
 is noisy. 
In the opposite scenario, where Bob and Charu are the senders,  we find that the dense coding capacity also follow the usual monogamy relation of 
quantum correlations.
We subsequently  proved that a strict monogamy holds for the case when there are an arbitrary number of senders and a single receiver in arbitrary dimensions. 
This has potential applications in quantum networks, involving several senders and several receivers.


\begin{thebibliography}{999}
\bibitem{dharma-versus-giraffe} Below, we also consider the case when the quantum channel that takes the Alice's part of the ensemble states \(\varrho^i_{AB}\) is noisy.
\end{thebibliography}
\end{document}